\documentclass{emulateapj}

\def\gsim{\;\rlap{\lower 2.5pt
 \hbox{$\sim$}}\raise 1.5pt\hbox{$>$}\;}
\def\lsim{\;\rlap{\lower 2.5pt
   \hbox{$\sim$}}\raise 1.5pt\hbox{$<$}\;}
\def\sfrunits{$M_\odot$ yr$^{-1}$ }
\def\microns{$\mu$m }
\def\micronsend{$\mu$m}
\def\ujy{$\mu$Jy}
\def\timeunits{$h^{-1}$ Gyr }

\def\lengthunits{$h^{-1}$ kpc }

\begin{document}

\title{Rest--Frame Ultraviolet to Near Infrared Observations of an Interacting Lyman Break Galaxy at $z = 4.42$}
\author{Joshua D. Younger\altaffilmark{1}, Jia--Sheng Huang\altaffilmark{1}, Giovanni G. Fazio\altaffilmark{1}, Thomas J. Cox\altaffilmark{1}, Kamson Lai\altaffilmark{1}, Philip F. Hopkins\altaffilmark{1}, Lars Hernquist\altaffilmark{1}, Casey J. Papovich\altaffilmark{2,3}, Luc Simard \altaffilmark{4}, Lihwai Lin\altaffilmark{5,6}, Yi--Wen Cheng\altaffilmark{7}, Haojin Yan\altaffilmark{8}, Du\v{s}an Kere\v{s}\altaffilmark{1}, \& Alice E. Shapley\altaffilmark{9}}
\altaffiltext{1}{Harvard--Smithsonian Center for Astrophysics, 60 Garden Street, Cambridge, MA 02138, USA}
\altaffiltext{2}{Steward Observatory, University of Arizona, 933 North Cherry Avenue, Tucson, AZ 85721, USA}
\altaffiltext{3}{{\it Spitzer} Fellow}
\altaffiltext{4}{Herzberg Institute of Astrophysics, National Research Council of Canada, 5071 West Saanich Road, Victoria, BC V9E 2E7, Canada}
\altaffiltext{5}{UCO/Lick Observatory, University of California, 1156 High Street, Santa Cruz, CA 95064, USA}
\altaffiltext{6}{Department of Physics, National Taiwan University, No. 1, Sec. 4., Roosevelt Road, Taipei 106, Taiwan}
\altaffiltext{7}{Institute of Astronomy, National Central University, Chung-li, Taiwan}
\altaffiltext{8}{Carnegie Observatories, 813 Santa Barbara Street, Pasadena, California, 91101 USA}
\altaffiltext{9}{Department of Astrophysical Sciences, Princeton University,  Peyton Hall, Ivy Lane, Princeton, NJ 08544}

\begin{abstract}

We present the rest--frame ultraviolet through near infrared spectral energy distribution for an interacting Lyman break galaxy at a redshift $z=4.42$, the highest redshift merging system known with clearly resolved tidal features.  The two objects in this system -- HDF--G4 and its previously unidentified companion -- are both $B_{435}$ band dropouts, have similar $V_{606}-i_{775}$ and $i_{775}-z_{850}$ colors, and are separated by $1\arcsec$, which at $z=4.42$ corresponds to 7 kpc projected nuclear separation; all indicative of an interacting system.  Fits to stellar population models indicate a stellar mass of $M_\star = 2.6\times 10^{10} M_\odot$, age of $\tau_\star = 720$ My, and exponential star formation history with an $e$--folding time $\tau_0 = 440$ My.  Using these derived stellar populations as constraints, we model the HDF--G4 system using hydrodynamical simulations, and find that it will likely evolve into a quasar by $z\sim3.5$, and a quiesent, compact spheroid by $z\sim 2.5$ similar to those observed at $z\gsim2$.  And, the existence of such an object supports galaxy formation models in which major mergers drive the high redshift buildup of spheroids and black holes.

\end{abstract}


\section{Introduction}

Recently, there have been great advances in understanding the nature and evolution of high--redshift galaxies.  Through either color--selection criteria \citep{steidel2003,franx2003,daddi2004}, Lyman--$\alpha$ emission \citep{rhoads2001, malhotra2002,ajiki2003,hu2004,taniguchi2005} or blank--field submillimeter surveys combined with radio observations \citep{smail2000,barger2000, ivison2002,borys2003,chapman2005,coppin2005} observers have compiled large catalogs of $z \gsim 2$ galaxies.  Subsequent photometric and spectroscopic followup has been successful at constraining the physical properties of these galaxy populations, including the evolution of the star formation rate (SFR) and stellar mass density out to $z \sim 3$ \citep{shapley2003,barmby2004,shapley2005,lai2007}.

One method for selecting high--redshift galaxies -- the Lyman break dropout techinque \citep{steidel1993} -- uses the $912\textrm{\AA}$ break to select $z \sim 3$ galaxies.  Spectroscopic followup has provided substantial samples of confirmed $z \sim 3-4$ galaxies \citep{steidel1999,steidel2003}.  These Lyman break galaxies (LBGs) are known to have high star formation rates relative to local galaxies of the same stellar mass, as revealed by their high rest frame UV luminosity \citep{steidel2003}.  Until recently, stellar masses for these objects were estimated using ground--based optical photometry.  At these wavelengths in the rest--frame, the luminosity is dominated by recent star formation, rather than the accumulated mass from older stellar populations.  Studies using the Infrared Array Camera \citep[IRAC:][]{fazio2004} on the {\it Spitzer} Space Telescope, which provides photometry out to $8\mu$m in the observed frame, have used the rest--frame K band luminosity to place more robust constraints on the bulk of the stellar content of LBGs at $z \sim 3$ \citep{rigopoulou2006}.

At the same time, recent theoretical modeling \citep{hopkinshernquist2006a,hopkinshernquist2006b,hopkins2007b,hopkins2007a} and simulations \citep{coxdutta2006} have suggested a link between merging galaxy populations, quasars, and present day ellipticals through the self--regulated growth of supermassive black holes (SMBHs) in gas--rich major mergers \citep{DiMatteo2005}.  The presence of large amounts of dust \citep{sawicki1998,calzetti2001,takeuchi2004} that implies a high SFR, in combination with clustering arguments \citep{adelberger2005}, and a merging fraction of $\sim 10-25\%$  at high redshift \citep{lotz2006} have suggested empirically that high--redshift LBGs may be the progenitors of those same present--day ellipticals \citep{pettini1998}.   As a result, major mergers involving LBGs at high redshift may provide observational evidence of a link between these populations.

In this work we present optical through infrared observations of HDF--G4, an LBG with a spectroscopically confirmed redshift of $z = 4.42$ in the Hubble Deep Field North (HDFN) with $\alpha(\textrm{J2000}) =$ 12:37:20.57 and $\delta(\textrm{J2000}) = $+62:11:6.08 from the catalog of \citet{steidel1999}.  This object, and its previously unidentified companion, constitute the most distant  interacting system observed to date for which the resolved morphology clearly suggests a merger. In \S~\ref{sec:phot}, we present multiwavelength data for the HDF--G4 system.  From these data, in \S~\ref{sec:pops}, we use stellar population synthesis models to derive stellar masses and star formation rates (SFRs) for each component.  In \S~\ref{sec:sims}, we use the observed stellar populations and morpholgy to constrain a model of the HDF--G4 system using hydrodynamical simulations.  Finally, in \S~\ref{sec:discuss}, we discuss the implications of the predictions of our model for the future evolution of the HDF--G4 system.  Our analysis suggests that HDF--G4 will be a high--redshift passive spheroid by $z\sim2.5$, similar to those observed by \citet{labbe2005} and \citet{zirm2007}.  For the proceeding analysis, we assume the concordance cosmological model; a flat $\Lambda$CDM cosmology with $\Omega_\Lambda = 0.7$ and $h=0.7$.  All magnitudes presented are in the AB system.

\section{Multi-Wavelength Photometry of HDF-G4}
\label{sec:phot}
The Great Observatories Origins Deep Survey (GOODS), which includes the HDFN, is the deepest 
multi-wavelength survey, detecting galaxies at extremely high redshifts
in optical and IR bands \citep{mobasher2005,yan2006,lai2007}, and potentially interacting systems at very high redshift at $z \gsim 4$ \citep{rhoads2005,yan2005}.   The wealth of multi--wavelength data available for GOODS fields, from the X--Ray to radio, makes it ideal for studying the properties of such interesting systems.

We visually inspected the available optical imaging data, and find that HDF-G4 appears to be one object in the ground--based Subaru images \citep{capak2004}, but is resolved into two objects in the ACS mosaic \citep{giavalisco2004}. Aperture photometry obtained with {\sc Sextractor}, a publicly available source detection and photometry package \citep{bertin1996}, for both the primary (C1) and secondary (C2)  components were taken from the public catalog for all four ACS mosaics: F435W ($B_{435}$), F606W ($V_{606}$), F775W ($i_{775}$), and F850LP ($z_{850}$).  C2 is 1 magnitude fainter in the ACS $i_{775}$ band, and $\sim1\arcsec$ away from C1.   Furthermore, the position angle and center of the slit--mask indicate that the observed optical spectrum is of C1 only.  

\begin{table*}
\setlength{\tabcolsep}{0.03in.}
\caption{Photometry of Different Components of the HDF--G4 System}
{\scriptsize
\begin{tabular}{ccccccccccc}
\hline
\hline
& $B_{435}$ & $V_{606}$ & $i_{775}$ & $z_{850}$ & $J$ & $K_s$ & 3.6\micron & 4.5\micron & 5.8\micron & 8.0\micron \\
& [mag] & [mag] & [mag] & [mag] & [mag] & [mag] & [\ujy] & [\ujy] & [\ujy] & [\ujy] \\ 
\hline
C1 & $>27.2$ & $25.87\pm0.02$ & $24.72\pm0.02$ & $24.77\pm0.02$ & $23.65\pm0.59$ & $23.59\pm 0.18$ & $1.34\pm 0.22$ & $0.92\pm 0.34$ & $0.68\pm 3.47$ & $0.79\pm 0.94$ \\
C2 & $>27.2$ & $26.43\pm 0.03$ & $25.67\pm0.03$ & $25.80\pm 0.04$ & \nodata & \nodata & \nodata & \nodata & \nodata & \nodata \\
Bridge$^a$ & $>27.74$ & $27.92\pm0.19$ & $26.78\pm0.12$ & $26.72\pm0.15$ & \nodata & \nodata & \nodata & \nodata & \nodata & \nodata \\
\hline
\hline
\tablenotetext{a}{Errors and magnitude limits estimated using a Monte--Carlo analysis.\\}
\label{tab:photometry}
\end{tabular}
}
\end{table*}

\begin{figure}
\plotone{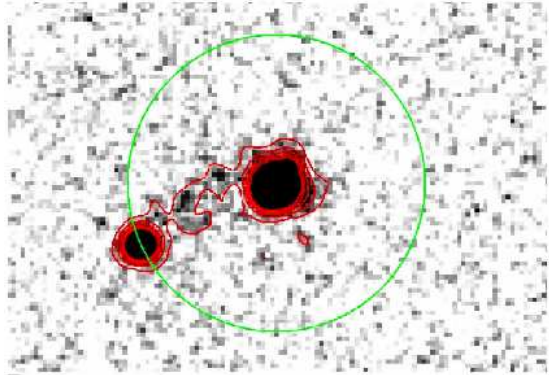}
\caption{Stacked optical imaging data centered on HDF--G4 for all four ACS filters: $B_{435}$, $V_{606}$, $i_{775}$, and $z_{850}$.  Red contours correspond to 2--5 standard deviations above the noise.  The green circle has a $1\arcsec$ radius, which at this redshift corresponds to $\sim 7$ kpc in our assumed cosmology.  Note the material exchanged between the two sources, contained within a $2$--$\sigma$ contour.  This is, despite its low per--pixel surface brightness, a highly statistically significant feature that indicates an interacting system.}
\label{fig:opticalimg}
\end{figure}

  There are three pieces of evidence suggesting that C1 and C2 are physically associated with each other. The first is that both objects are $B_{435}$ dropouts with similar observed optical colors, indicating that they are at comparable redshift.   The second is their relative angular separation, which at $z=4.42$ corresponds to a projected internuclear distance of 7 kpc.  This is similar to local interacting systems such as the ``Antennae'' \citep{whitmore1995}, II ZW 96 \citep{goldader1997}, or ultraluminous infrared galaxies \citep[ULIRGs;][]{rigopoulou1999,surace2000}.  The third, and most compelling piece of evidence is a material bridge connecting both components. This feature is low surface brightness -- typically no more than 2$\sigma$ above the sky noise per pixel -- but highly statistically significant; a Monte--Carlo analysis indicates that it is detected at 10$\sigma$ in the stacked optical image (see Figure~\ref{fig:opticalimg}).  It furthermore has similar colors to C1 and C2, and is not detected in $B_{435}$ (see Table~\ref{tab:photometry}).  

    This system is also marginally detected in the CFHT WIRCam J and K band images (Lin 2006, Simard 2006, private communication).  The centroid of the K-band counterpart is $0.35\arcsec$ offset from C1, 
but $0.8\arcsec$ from C2.  Given the $0.9\arcsec$ seeing in the J and K images, we argue the detected J- and K-band flux is dominated by C1. 

\begin{figure}
\plotone{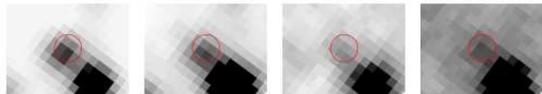} 
\caption{Infrared imaging data, with negative contrast, centered on HDF--G4, in all four IRAC channels: (left to right) 3.6, 4.5, 5.8, and 8.0 $\mu$m.  The red circles have a $1\arcsec$ radius.  HDF--G4 is clearly detected at 3.6\microns  and 4.5\microns, and marginally detected at 5.8\microns  and 8.0\microns.}
\label{fig:irimg}
\end{figure}

 The GOODS IRAC image is very deep, and thus both C1 and C2 should be detected
at least at 3.6 and 4.5\micron. Indeed, we find significant detections at 3.6 and 4.5\micronsend, and marginal detections at 5.8 and 8.0\microns (see Figure~\ref{fig:irimg}).  We obtain aperture photometry (see Table~\ref{tab:photometry}) for HDF--G4, after subtracting out bright neighbors to minimize contamination using {\sc StarFinder}, a publically available PSF--fitting photometry package \citep{diolaiti2000}.  Errors were estimated using a Monte--Carlo analysis.  IRAC, however, has $2.1\arcsec$
resolution, and thus cannot resolve C1 and C2. We use the $z_{850}$ flux ratio to estimate the C2 contribution to the total IRAC flux of this system. C2 has a somewhat bluer color with 
$V_{606}-z_{850}=0.64$, as compared to C1 with $V_{606}-z_{850}=1.09$. This is very 
similar to a merging system at z=3.01 found by \citet{huang2006}, in which the
IR flux densities are dominated by the red luminous component. We argue
that C2 should also have bluer $z_{850}-[3.6]$ color than C1, and use the $z_{850}$ flux ratio to estimate an upper limit of 10\% on the contribution of C2 to the unresolved IRAC flux densities.  

 The HDFN was additionally covered by the Very Large Array at $1.4$ GHz to a depth of 40 $\mu $Jy \citep{richards2000}, MIPS to a depth of $\sim 70 \mu$Jy at 24\microns, and the {\it Chandra} X--Ray Telescope to depths of $2.5\times 10^{-17}$ and $1.4 \times 10^{-16}$ erg cm$^{-2}$ in the soft (0.5-2.0 keV) and hard (2-10 keV) bands respectively \citep{alexander2003}.  There was, however, no detection of either HDF--G4 or its companion in the publicly released source lists or imaging data.  This is all consistent -- and in particular the lack of 24\microns or 1.4 GHz emission -- with the high redshift of HDF--G4.  

\begin{figure}
\plotone{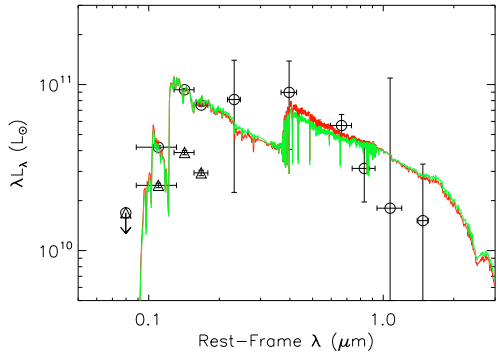}
\caption{The SED of HDF--G4 (open circles) and its companion (open triangles) from the rest--frame UV to the NIR, along with the best--fit \citet{bruzual2003} stellar population synthesis models for an exponential (red; ESF) and continuous (green; CSF) star formation history.  IRAC data for 5.8\microns and 8.0\microns are shown for completeness, but are excluded from the fitting because of their low significance.}
\label{fig:sed}
\end{figure}

We present the spectral energy distribution (SED) of HDF--G4 (open circles) and and its companion (open triangles), along with best fit models (see \S~\ref{sec:pops}) in Figure~\ref{fig:sed}.  All the available photometry for C1, C2, and the bridge are presented in Table~\ref{tab:photometry}.

\section{Analysis of the Stellar Populations}
\label{sec:pops}

\begin{figure*}
\plotone{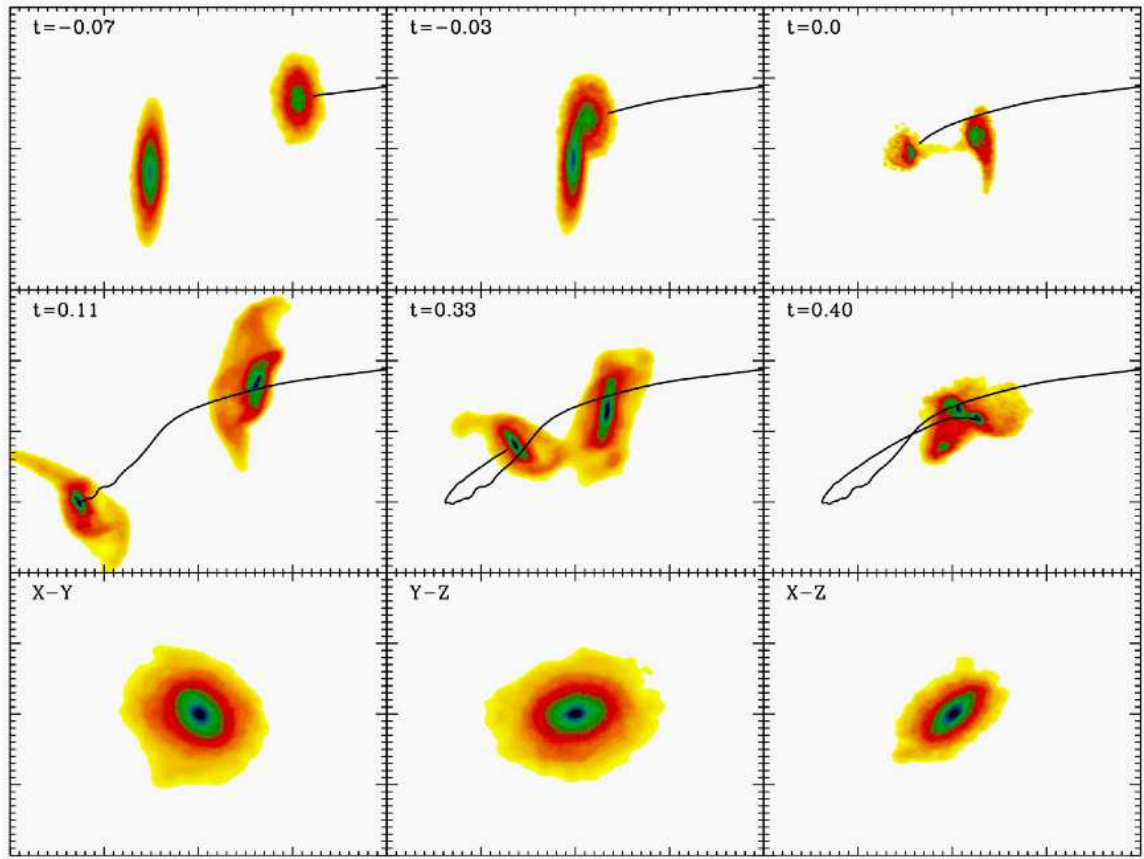}
\caption{Schematic of the merger simulation that best matches our observations.  The top six panels show the stellar surface mass density looking down on the orbital plane.  They are 60 \lengthunits on a side , the time relative to the start of the simulation in \timeunits is given in the upper right hand corner, and the black curve shows the path of the center of the smaller component (C2).  The bottom three panels show the remnant spheroid viewed from three different projections.}
\label{fig:panels}
\end{figure*}

Our observations of each component of the HDF--G4 system can be used to constrain the underlying stellar population at the time of our observations.  To estimate the age and mass of the stellar population of the primary component of HDF--G4 (C1), we fit its SED to a grid of population synthesis models \citep{bruzual2003}, assuming a \citet{salpeter1955} IMF, with three different parameterized star formation histories: a single stellar population (SSP) instantaneous burst, continuous star formation (CSF), and an exponential decay $\tau$--model (ESF).  We consider solar and sub--solar ($Z_\odot/200$) models, in addition to including the effects of \citet{calzetti2000} dust extinction and \citet{madau1995} Lyman--$\alpha$ forest absorption.  Our fits exclude observations in IRAC Channels 3 and 4, which are poorly constrained by the measurement.  

Results for the best--fit CSF and ESF models are shown in Figure~\ref{fig:sed};  the SSP models were excluded because they were both a poor fit to the data.  Requiring the age of the stellar population $\tau_\star$ to be less than a Hubble time at $z=4.42$ further excludes the sub--solar CSF and ESF models.  The best fit overall, with $\chi^2/\textrm{d.o.f.} = 1.46$ was the solar metallicity ESF model, with $e$--folding time $\tau_0 = 440$ My, stellar age $\tau_\star = 720$ Myr, stellar mass $M_\star = 2.6\times 10^{10} M_\odot$,  and dust extinction $E(B-V) = 0$.  The best--fit CSF model, with $\chi^2/\textrm{d.o.f.} = 2.24$, had $\tau_\star = 1020$ Myr, $M_\star = 2.6\times 10^{10} M_\odot$, $E(B-V) = 0.05$, and star formation rate SFR = 32 \sfrunits.  These results are somewhat sensitive to the choice of IMF; e.g., assuming a \citet{chabrier2003} distribution will reduce the inferred stellar masses by $\sim30-40\%$.  These inferred stellar masses and SFRs are broadly consistent with studies applying a similar analysis to large populations of LBGs \citep{papovich2001,shapley2001}, and with 8\microns selected LBGs with similar infrared--to--optical colors at $z\sim3$ \citep{rigopoulou2006}.  We note that our best fit stellar populations indicate minimal dust extinction, which is unusual for LBGs but is consistent with the least obscured objects at $z\sim2-3$ \citep{papovich2001,shapley2001}.

We do not have sufficient spectral coverage to fit models to the photometry for C2 or the bridge.  However, at $z=4.42$, the observed optical images cover the rest--frame ultraviolet continuum from 800-2310\AA.  The integrated spectrum in this waveband is dominated by young stars, and therefore scales linearly with the star formation rate \citep{kennicutt1998a,madau1998}.  Using the calibrations of \citet{kennicutt1998a} for the intrinsic luminosity at 1500 \AA, which assume CSF and a \citet{salpeter1955} IMF, we find that our $i_{775}$ (rest--frame $\lambda_{eff} \approx 1421$\AA) photometry implies SFR $\approx$ 9 \sfrunits.  Similarly, the bridge traces significant off--nuclear star formation, with SFR $\approx$ 3 \sfrunits as inferred from the $i_{775}$ magnitude.  If a \citet{scalo1986} IMF is used, these estimates increase by a factor of $\sim 2$.  

The lack of a detection of HDF--G4 in deep {\it Chandra} observations can constrain the relative contributions of an active galactic nucleus (AGN) and stellar components of the bolometric luminosity in C1.  From the AGN SED template of \citet{hopkinsrichards2007}, we find that the limiting flux of the hard band  ($2-8$ keV) puts a strict upper limit, in the absence of significant scattering or absorption by dust, of $L_{AGN} \lsim 2 \times10^{10} L_\odot$ on the bolometric AGN luminosity.  We believe that neglecting the effects of dust on the emitted X--rays is a good approximation for two reasons: the population synthesis fits indicate that there is little dust attenuation in the optical, and at $z=4.42$ the {\it Chandra} hard band probes very hard X--rays ($11-44$ keV) that are largely unaffected by dust in the line of sight.  The integrated intrinsic stellar luminosity, from the best--fit ESF model, is $L_{stars} = 2.0\times 10^{11} L_\odot$, which constrains the ratio of the luminosity of the AGN to the stellar component of HDF--G4 to $L_{AGN}/L_{stars} \lsim 0.1$; C1 is clearly starburst dominated.

\section{Comparison to Hydrodynamical Simulation}
\label{sec:sims}

In order to gain some insight into the nature of the HDF--G4 system, we have employed hydrodynamic simulations and designed a model for the encounter.  Our simulations were preformed using an updated version of the publicly available N--Body/SPH (Smoothed Particle Hydrodynamics) code {\sc Gadget2} \citep{springel2005}.  They include star formation, supernova feedback, and black hole accretion; for a detailed description of our methodology, see \citet{springeldimatteo2005b,springel2005b}.  Using this code, we track the interaction and merger of two disk galaxies whose properties are scaled, as in \citet{robertson2006a}, to be appropriate for a redshift of $z=5$.

The progenitor galaxy models are motivated by the stellar mass and size of the observed components of the HDF--G4 system.  In particular, the observed relative fluxes of C1 and C2, assuming a constant $M/L$ ratio, imply a 2:1 interaction.  This is reproduced with progenitor disks initialized with circular velocities of $V_{200} = 320$ and 200 km s$^{-1}$ respectively, where $V_{200}$ is the Keplerian circular velocity at a mean overdensity of 200$\rho_c$.  Both models have baryon fractions of 5\% and initial gas fractions of $f_g = 0.8$, which is consistent with gas fractions observed in $z\gsim2$ disks \citep{erb2006a}, and is further motivated by models \citep{hopkins2007d} to explain large black hole to host stellar mass ratios at $z\gsim 2$ \citep{peng2006}.  They are realized with $10^6$ and $5\times 10^5$ dark matter, and $1.2\times 10^5$ and $6\times 10^4$ baryonic particles respectively.

Our high redshift merger model assumes that HDF--G4 is witnessed soon after the first passage of the two galaxies.  This is based on the appreciable projected physical separation that implies a lower limit of 7 kpc on the internuclear distance, regular appearance, and bridge of material that connects the two components.  The bridge in particular, while potentially reflective of the irregular rest--frame UV morphology of star--forming systems \citep[e.g.,][]{law2007},  is also a characteristic feature of galaxy interactions at low redshift and provides the strongest constraints on the mutual orbit.  A prograde encounter for either galaxy can be ruled out based upon the absence of the symmetric tidal features that these interactions produce.  More likely is a fairly direct encounter in which the galaxies interpenetrate at first passage and material is strewn out to form a bridge between them.  After several trials, the best fitting orbit was parabolic -- in agreement with likely orbital geometrics from cosmological simulations \citep{benson2005,khochfar2006} -- with a small perigalacticon of 1 kpc and both disks oriented nearly perpendicular to the orbital plane.

\begin{figure}
\plotone{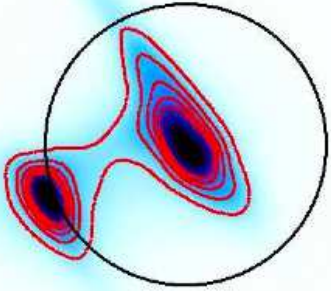}
\caption{The rest--frame $1500\textrm{\AA}$ surface brightness of the same snapshot in Figure~\ref{fig:panels}, from the viewing angle that best matches the HDF--G4 system.  This corresponds roughly to the observed $i$--band at $z=4.42$.  This image has the same scale and resolution as Figure~\ref{fig:opticalimg}, and has been convolved with the ACS PSF.  The black circle, as in Figure~\ref{fig:opticalimg}, represents 1\arcsec or 7 kpc projected physical separation, and the red contours show the isophotal shapes.  Note that the simulations are able to roughly reproduce the rest--frame morphology of HDF--G4, with a bridge between the two components.}
\label{fig:uvmorph}
\end{figure}

\begin{figure}
\plotone{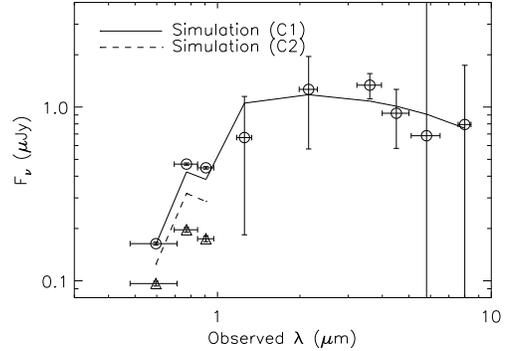}
\caption{Simulated fluxes at $z=4.42$ versus our observations of the HDF--G4 system for both components  over the physical scales in the ACS image.  The squares and solid line are the observed and simulated SED for C1, and the triangles and dashed line are the observed and simulated SED for C2.}
\label{fig:merger_colors}
\end{figure}

Snapshots of the stellar mass distribution, including three different projections of the remnant, at different stages of the encounter are shown in Figure~\ref{fig:panels}.  We treat the stellar particles as \citet{bruzual2003} SSPs with ages and metallicities according to their formation time.   Stars initialized with the progenitor disks are assumed to have formed 0.7 Gyr before the start of the simulation with low metallicity $Z=1\times10^{-5}$.  We neglect absorption and scattering by dust because the stellar population synthesis analysis in \S~\ref{sec:pops} indicates that HDF--G4 is virtually dust--free.  The best match to the observed optical morphology (rest--frame UV $\lambda\sim1500\textrm{\AA}$) of HDF--G4 was at $t=0.17$ Gyr after the start of the simulation, and is shown in Figure~\ref{fig:uvmorph}.  The snapshot shows the rest--frame $L_{1500\AA}$ surface brightness at the resolution of the observations in Figure~\ref{fig:opticalimg} for $z=4.42$, and convolved with the ACS PSF.  The stellar mass at this snapshot is $2.9\times10^{10}\, M_{\odot}$ with a total SFR of 59 \sfrunits; both within the observational constraints given the uncertainty in the IMF and relative $M/L$ ratios of C1 and C2.  The synthetic SEDs of the two components over the physical scales in the ACS image, shown in Figure~\ref{fig:merger_colors}, are also a close match to the observations.  Finally, the total  luminosity from accretion onto the black hole is $L_{AGN} = 8\times 10^8\, L_{\odot}$, which is well within the upper limit of $L_{AGN} < 2\times 10^{10} L_{\odot}$ imposed by the observations.  The ability of simulations to reproduce the observable features of this object supports our interpretation that HDF--G4 is an early--stage merger.

Finally, we note as a caveat that the isophotal shapes in Figure~\ref{fig:uvmorph} are somewhat more elongated than those in the optical imaging data.  Though our population synthesis modeling suggests that the overall effects of dust in the HDF--G4 system are small, because this is the rest--frame UV, even small amounts of dust in these elongations could substantially change their appearance in simulated observations.  The structure of these features is also sensitive to the specific orbital geometry and the detailed structure of the progenitor disks, both of which may not be precisely captured in our modeling.  The bridge, on the other hand, is produced by ram--pressure stripping of gas when the progenitors interpenetrate.  Furthermore, if there were small amounts of dust in the system, this stripping would not necessarily carry the dust along with it.  As a result, though the detailed isophotal shapes of the two components are sensitive to the initial conditions of our simulations, we believe that production of a UV--bright bridge between the two components is generic for any encounter with a small impact parameter, and is therefore the best evidence for an ongoing merger and the applicability of our modeling.

\section{Discussion}
\label{sec:discuss}

\begin{figure}
\plotone{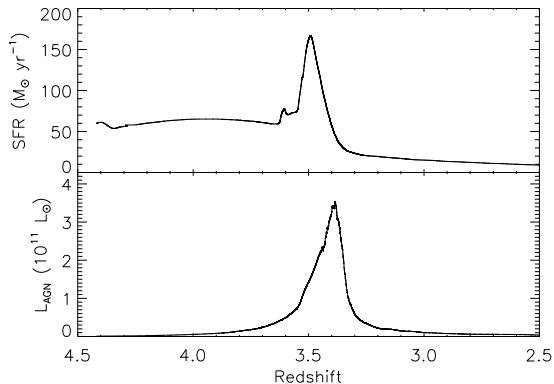}
\caption{The future evolution of the SFR (top panel) and the AGN luminosity (top panel) of the HDF--G4 system as predicted by our simulations.  This suggests that HDF--G4 will be a quasar at $z\approx 3.5$, and a quiescent spheroid at $z\approx 2.5$.}
\label{fig:evol}
\end{figure}

In Figure~\ref{fig:evol}, we use our simulation to predict the future evolution of the HDF--G4 system.  This indicates that the two components will finally coalesce, with a peak in the starburst and AGN luminosity, around a redshift of $z\approx 3.5$.  At this time, the HDF--G4 system will be a $L_{bol}\approx 4\times 10^{11}\, L_{\odot}$ quasar.  Based on the \citet{hopkinshernquist2006a,hopkinssommerville2006} quasar luminosity function models, we would expect there to be of order a couple such systems in the GOODS--North field.  Although we cannot robustly constrain the space densities of these mergers, it is conservatively consistent with the expected counts. 

Our simulations also suggest that by a redshift of $z\approx 2.5$, HDF--G4 will have a spheroidal morphology (see Figure~\ref{fig:panels}), a stellar mass of $M_\star = 7.4\times 10^{10}$ $M_{\odot}$, and a SFR of 8.8 \sfrunits; a slowly star--forming but not entirely passive elliptical.  Its colors will be red, with $(J-K)_{AB} \approx 1.2$, which is close to the color cut of $(J-K)_{AB} > 1.3$ for Distant Red Galaxies \citep[DRGs:][]{vandokkum2003}.  The remnant also has a small effective radius -- measured as in \citet{coxdutta2006} -- relative to local spheroids of the same mass, with $R_e = 1.4\pm0.17$ kpc and an average inner stellar surface mass density within $R_e$ of $\sigma_e=(0.6\pm 0.03)\times 10^{10}$ $M_{\odot}$ kpc$^{-2}$, where the errors take into account projection effects.  This makes it similar to the compact, quiescent high redshift spheroids at $z\sim2$ observed by \citet{labbe2005} and \citet{zirm2007}, who find effective radii of $R_e\sim 0.5-1.1$ kpc, inner stellar surface mass densities of $\sigma_e\sim 0.7-4$ $M_{\odot}$ kpc$^{-2}$, and stellar masses of $M_\star\sim5-9\times 10^{10}$ $M_{\odot}$.

The existence of such an object further lends observational support to galaxy formation models which argue that mergers dominate the high-redshift buildup of spheroids and black holes \citep{hopkinsrichards2007,croton2006}. It is difficult, on the other hand, to reconcile this with models which argue that spheroid and black hole formation at high redshift is driven by direct gas collapse/cooling \citep{granato2004} or disk instabilities \citep[e.g.,][]{bower2006}. In the latter model, for example, mergers contribute only $\sim0.1\%$ of bulge and BH growth at these redshifts, which would predict systems like HDF-G4 should be too rare to observe in small fields, should be already ``dry'' (i.e. spheroid-dominated, with much lower current SFRs), and should have pre-existing large BHs from their disk phase (given the disk masses and gas fractions necessarily present at these redshifts) which would, given gas is clearly still present (from the observed SFR), inescapably make the system a ($\sim 3\times10^{12}\,L_{\odot}$) quasar at the observed time.  Therefore, it will be interesting to see if future observations of wider fields reveal HDF--G4 to be anomalous, or representative of a typical, albeit brief, phase of LBG evolution.

\section{Conclusion}

We present the rest--frame UV through NIR SED for HDF--G4, an interacting Lyman Break galaxy in GOODS-North with a spectroscopically confirmed redshift of $z=4.42$ \citep{steidel1999}, using data obtained from ACS, WIRCam, and IRAC.  The two objects in this system -- HDF--G4 and its previously unidentified companion -- are both $B_{435}$ band dropouts, have similar $V_{606}-i_{775}$ and $i_{775}-z_{850}$ colors, and are separated by $1\arcsec$, which at $z=4.42$ corresponds to 7 kpc projected nuclear separation, and a bridge of material between them.  We apply stellar population synthesis models \citep{bruzual2003} to the SED of HDF--G4, and find a best--fit population with solar metallically, an exponential star formation history, with $e$--folding time $\tau_0 = 440$ My, stellar age $\tau_\star = 720$ Myr, stellar mass $M_\star = 2.6\times 10^{10} M_\odot$,  and dust extinction $E(B-V) = 0$ with a reduced $\chi^2/\textrm{d.o.f} = 1.46$.  The observed stellar population, combined with the SED, system morphology, and projected nuclear separation, are used to constrain a model of the HDF--G4 system using a hydrodynamical simulation.  This analysis suggests that HDF--G4 is the potential progenitor of a $z \approx 3.5$ quasar with $L\approx 4\times10^{11}$ $L_{\odot}$, and a compact ($R_e= 1.4$ kpc) quiescent $z\sim 2.5$ spheroid consistent with the population observed by \citet{labbe2005} and \citet{zirm2007}.  Furthermore, the existence of such an object supports galaxy formation models in which major mergers drive the high redshift buildup of spheroids and black holes \citep[e.g.,][]{hopkinsrichards2007}.

\acknowledgements

We thank the referee for their helpful comments.  This work is based on observations made with the {\it Spitzer} Space Telescope, which is operated by the Jet Propulsion Laboratory, California Institute of Technology, under NASA contact 1407.  CFHT observations are supported through the Taiwan CosPA project.


\end{document}